\documentclass[aps,prb,twocolumn,superscriptaddress]{revtex4-2}

\usepackage[utf8]{inputenc}
\usepackage{comment}
\usepackage{mathtools}
\usepackage{amsfonts}
\usepackage{mathrsfs}
\usepackage{physics}
\usepackage{slashed}
\usepackage{tensor}
\usepackage{dsfont}
\usepackage{bbm}
\usepackage{orcidlink}
\usepackage{graphicx}
\usepackage{color, float}
\usepackage{array}
\usepackage[abs]{overpic}

\usepackage{tikz}
\usetikzlibrary{decorations.pathmorphing}

\usepackage{placeins}

\usepackage{makecell}

\usepackage{xspace}
\usepackage{siunitx}
\usepackage{xfrac}
\usepackage{hyperref}
\usepackage[nameinlink]{cleveref}
\usepackage{appendix}

\usepackage{adjustbox}

\usepackage{xifthen}
\usepackage{xcolor}
\hypersetup{
	colorlinks,
	linkcolor={red!75!black},
	citecolor={blue!75!black},
	urlcolor={blue!75!black}
}

\newcommand{\gettitle}{Critical droplets and metastability  in long-range chains}

\newcommand{\getZuerichAffiliation}{\affiliation{Institut f{\"u}r Theoretische Physik, ETH Z{\"u}rich, Wolfgang-Pauli-Str. 27, 8093 Z{\"u}rich, Switzerland}}

\hypersetup{
	pdftitle={\gettitle},
	bookmarksopen=true,
	bookmarksopenlevel=2,
	bookmarksnumbered=true
}

\begin{document}
	
\title{\gettitle}

\author{Laura Batini \orcidlink{https://orcid.org/0009-0007-4129-4251}}
\thanks{L.B. and V.P. contributed equally to this work.\\
 E-mail L.B. at: lbatini@ethz.ch\\
 E-mail V.P. at: vpagni@ethz.ch. }
 \getZuerichAffiliation
\author{Valerio Pagni \orcidlink{https://orcid.org/0009-0007-5323-4651}}
\thanks{L.B. and V.P. contributed equally to this work.\\
 E-mail L.B. at: lbatini@ethz.ch\\
 E-mail V.P. at: vpagni@ethz.ch. }
\getZuerichAffiliation
\author{Nicolò Defenu \orcidlink{https://orcid.org/0000-0002-3401-3665}}\getZuerichAffiliation
\affiliation{CNR-INO, Area Science Park, Basovizza, I-34149 Trieste, Italy}

\begin{abstract} 
Metastable decay in first-order phase transitions proceeds through the nucleation of critical droplets. In the one-dimensional nearest-neighbor Ising chain, the cost of the two walls bounding a droplet is independent of its size. For interactions decaying as $1/r^{1+\sigma}$, this surface cost instead depends on the droplet size and competes with the bulk free-energy gain. The critical radius scales as $R_c\propto h^{-1/\sigma}$ at small $h$, where $h>0$ is the bias field.
The value $\sigma=1$ separates two regimes: for $\sigma<1$, the activation barrier diverges as $h\to0^+$; at $\sigma=1$, it grows logarithmically; and for $\sigma>1$, it remains finite, as in the nearest-neighbor chain. A nonlocal $\phi^4$ field theory yields critical droplets with algebraic tails and confirms the predicted radius and barrier scalings, while stochastic long-range dynamics independently reproduces the critical-radius scaling. 
\end{abstract}

\maketitle

\section{Introduction}

Metastability and nucleation are central to the dynamics of first-order transitions in systems ranging from condensed matter to quantum field theory~\cite{coleman1977fate, Lagnese:2021grb,Batini:2023zpi,Lagnese:2023xjg,Batini:2024zst}. A system can remain trapped in a locally stable state
even when a lower-free-energy phase is available, and its decay then
proceeds through rare fluctuations \cite{Langer:1969bc}. In the nucleation picture,
such a fluctuation creates a localized domain, or droplet, of the stable
phase within the metastable background \cite{PhysRevB.60.14525,jacquier2025exploring}. Small droplets shrink and disappear,
whereas sufficiently large ones grow and eventually convert the surrounding
system. The critical droplet separating these two regimes sets the activation barrier for nucleation.

How this picture changes when interactions extend over long distances remains comparatively less understood. Trapped-ion simulators realize tunable power-law Ising couplings~\cite{monroe2021programmable,britton2012engineered} and have recently probed bubble-nucleation dynamics~\cite{luo2025quantum}, and false-vacuum decay has also been investigated in Rydberg arrays~\cite{browaeys2020many,Chao:2025rhr}. Photon-mediated interactions in cavity-coupled gases provide a complementary setting for metastability and switching~\cite{ritsch2013cold,Hruby_2018,Baumgartner:2024cmy}, while ultracold-gas experiments directly probe thermal effects in metastable decay~\cite{zenesini2024observation,Cominotti:2025qia}. These developments motivate understanding how interaction range affects thermally activated nucleation~\cite{Langer:1969bc,affleck1981quantum,linde1981fate}, a decay mechanism distinct from the quantum tunneling studied for long-range systems~\cite{Pagni:2026biq}.

  \begin{figure}
\centering
\includegraphics[width=.49\linewidth]{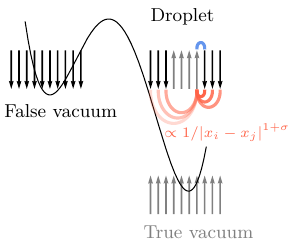}
\includegraphics[width=.49\linewidth]{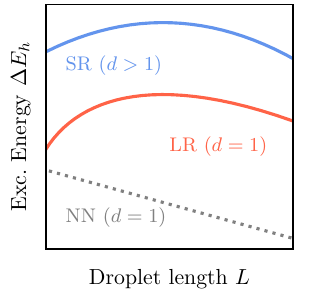}
\caption{Long-range interactions produce a mesoscopic critical droplet in the one-dimensional lattice chain. Left: a stable-phase domain embedded in the metastable background, with power-law couplings connecting distant spins. Right: schematic excitation energy $\Delta E_h(L)$ for the one-dimensional nearest-neighbor (NN) and long-range (LR) chains, compared with a higher-dimensional short-range (SR) case. }
 \label{fig:sketch}
 \end{figure}
In the thermal regime, false-vacuum decay is naturally described as a stochastic escape process. At a coarse-grained level, the dynamics of a non-conserved scalar order parameter can be described by relaxational dynamics~\cite{hohenberg1977theory,cardy1996scaling,tauber2014critical,ihssen2025nonperturbative,pagni2026critical,Batini:2026ugg,alford1993thermal}
\begin{equation}
    \partial_t\phi(x,t)
    =
    -\frac{\delta F[\phi]}{\delta\phi(x,t)}
    +\xi(x,t),
    \label{eq:langevin}
\end{equation}
where $F[\phi]$ is the free-energy functional and $\xi$ represents a zero-mean Gaussian noise, whose strength is fixed by temperature through the
fluctuation--dissipation relation: $\left\langle\xi(x, t) \xi(x^{\prime}, t^{\prime})\right\rangle=2 T \delta(x-x^{\prime}) \delta(t-t^{\prime})$. In the weak-noise regime, the associated
escape problem is described by Langer's theory of thermally activated
nucleation \cite{Langer:1969bc,affleck1981quantum, kamenev2005manybodytheorynonequilibriumsystems,gould2021effective,hirvonen2026langer}. For a metastable homogeneous configuration
$\phi_{\rm FV}$ and a lower-free-energy stable configuration
$\phi_{\rm TV}$, decay is controlled by a critical configuration
$\phi_c(x)$, whose excess free energy
$
    \Delta F_c=F[\phi_c]-F[\phi_{\rm FV}]
$
sets the activation barrier. To leading exponential accuracy, Langer's theory gives
$
    \tau_{\rm nuc}\simeq\tau_0\,e^{\Delta F_c/T}
$ in the limit where noise is weak relative to the activation barrier.

A central assumption underlying the conventional nucleation picture is
locality \cite{coleman1977fate}. 
For short-range interactions, the energetic cost of creating a domain is concentrated near its interface, while the free-energy gain arises from its bulk~\cite{binder1987theory}. Their competition produces the familiar critical
droplet. With long-range interactions, distant parts of the domain remain
directly coupled, so the energetic cost of a droplet depends on its overall
size and shape \cite{mukamel2009notesstatisticalmechanicssystems}.

The effect is clearest in one dimension. In the nearest-neighbor Ising chain, a stable-phase domain is bounded by two point-like walls whose energy is independent of their separation. Enlarging an existing domain therefore only increases the bulk energy gain: the energetic bottleneck is wall-pair creation, with no mesoscopic maximum along the contiguous-droplet path. For power-law couplings $1/r^{1+\sigma}$, the wall-pair energy grows with separation, and enlarging the domain by one site costs an energy that decays only as $L^{-\sigma}$. Its competition with the bulk gain produces a critical droplet for every $\sigma>0$, as illustrated in Fig.~\ref{fig:sketch}. For $\sigma<1$ the pair energy grows without bound, as $L^{1-\sigma}$, so that the two walls are confined, reminiscent of the confinement of quark--antiquark pairs and of the confined domain-wall excitations of long-range quantum spin chains~\cite{liu2019confined}. 
This behavior resembles that of short-range systems in $d>1$, where the interfacial free energy scales as $L^{d-1}$, as illustrated by the blue curve in Fig.~\ref{fig:sketch}. 

Rigorous lattice studies of the one-dimensional long-range Ising chain have characterized its critical configurations~\cite{van2019nucleation,jacquier2025exploring}. More recently, droplet arguments and simulations of a two-dimensional probabilistic cellular automaton showed that weak long-range interactions can produce apparent bistability through a rapidly growing critical-droplet size~\cite{lazarides2025apparent}.

Building on these lattice results, we determine critical-droplet profiles and their excess free energies in a nonlocal continuum field theory. While on the lattice a droplet has sharp boundaries, the continuum theory describes smooth interfaces with algebraic tails extending into the metastable background. Its radius scales as $R_c\propto h^{-1/\sigma}$ at small $h$, where $h$ is the bias field favoring the stable phase. While its exponent varies smoothly across $\sigma=1$, the energy barrier that suppresses the decay of the metastable state behaves differently: as $h\to0^+$, it diverges for $0<\sigma<1$, grows logarithmically at $\sigma=1$, and approaches a finite value for $\sigma>1$. This change between divergent and finite barriers coincides with the interaction-range threshold for finite-temperature ordering in the long-range Ising chain~\cite{Dyson:1968up,thouless1969long,frohlich1982phase}.

We independently test the radius scaling using stochastic majority-voting dynamics on a microscopic long-range chain. Starting from droplets of controlled size, we locate the boundary between shrinking and growing configurations and recover the predicted scaling behavior. The lattice energetics, the field theory and the dynamics give the same radius scaling.

\section{Microscopic droplet energetics} \label{Sec2:microscopic droplet energetics}

We begin with the lattice model, where the energy of a droplet can be
written explicitly. The Hamiltonian of the Ising model in the presence of long-range couplings is \cite{Dyson:1968up}
\begin{equation}\label{eq:Ising_Hamiltonian}
    H[\boldsymbol{s}] = -J_0 \sum_{i<j} \frac{s_i s_j  }{|i-j|^{1+\sigma}} - h \sum_i s_i,
\end{equation}
where $s_i=\pm1$, $i,j$ label the lattice sites, and $J_0>0$.
We consider $\sigma>0$, for which the power-law interaction $J(r)\sim r^{-(1+\sigma)}$ is summable and the thermodynamic limit is well defined without Kac rescaling. This excludes the strongly long-range, non-additive regime $\sigma\leq0$, whose spectral and thermodynamic properties are qualitatively different~\cite{campa2009statistical,defenu2021metastability, defenu2023long, solfanelli2024long,kastner2026long}.
Without loss of generality we consider a positive bias $h>0$ favoring the $s_i=+1$ phase. The all-$+1$
configuration is then the unique ground state, while for sufficiently
weak $h$ the all-$-1$ configuration remains a local energy minimum and
provides the metastable background. 

On the lattice, the relevant droplet is a contiguous block of $L$ spins
in the stable phase embedded in the metastable background
\cite{van2019nucleation, mukamel2009notesstatisticalmechanicssystems}. We denote this configuration by $s_L$, with
$s_i=+1$ inside the block and $s_i=-1$ outside. To isolate the effect of long-range interactions, it is useful to first
consider the strictly short-range model obtained by retaining only
nearest-neighbor couplings. Creating an interval of length $L$ then
introduces two domain walls with total energy $4J_0$, independent of their
separation. Its excitation energy is
\begin{equation}
    \Delta E_h^{\rm SR}(L)=4J_0-2hL .
\end{equation}
Once the two walls have been created, increasing their separation always
lowers the energy, see the gray dotted line in Fig.~\ref{fig:sketch}. Hence, the nearest-neighbor Ising chain has no nontrivial
mesoscopic critical size: the only energetic bottleneck is the creation
of the two domain walls. This is a property of sharp lattice walls. In a
short-range continuum $\phi^4$ theory, kink and antikink attract through
their exponentially decaying tails, and a critical droplet exists, but
its size grows only logarithmically, $R_c\propto\ln(1/h)$~\cite{alford1993thermal,Pirvu:2024nbe}.
Long-range interactions instead produce the power-law growth derived below,
by making the energy of the two domain walls depend on their separation.
Relative to the uniform metastable state, only bonds connecting a spin inside the droplet to one outside it change their energy. Given a bond length $r$, their number is
$2\min(r,L)$, and each such bond costs
$2J_0/r^{1+\sigma}$. The excitation energy is therefore \cite{mukamel2009notesstatisticalmechanicssystems,liu2019confined}
\begin{equation}
\begin{split}
    \Delta E_h(L)
    &\equiv H[s_L]-H[s_{\rm meta}]
    \\
    &=
    4J_0
    \left[
        \sum_{r=1}^{L}\frac{1}{r^\sigma}
        +
        L\sum_{r=L+1}^{\infty}
        \frac{1}{r^{1+\sigma}}
    \right]
    -2hL.
    \label{eq:lattice_droplet_energy}
\end{split}
\end{equation}
For large $L$, its interaction contribution behaves as
\begin{equation}
\frac{\Delta E_{\rm int}(L)}{4J_0}\simeq
\begin{cases}
\dfrac{L^{1-\sigma}}{\sigma(1-\sigma)},
& 0<\sigma<1,
\\
\ln L,
& \sigma=1,
\\
\zeta(\sigma)
-\dfrac{L^{1-\sigma}}{\sigma(\sigma-1)},
& \sigma>1 ,
\end{cases}
\label{eq:lattice_wall_energy}
\end{equation}
where $\zeta$ denotes the Riemann zeta function.
Thus, the interaction energy of a widely separated domain-wall pair
diverges for $\sigma\leq1$, whereas it approaches a finite value for
$\sigma>1$, already signaling a qualitative distinction at $\sigma=1$.
However, the
\emph{incremental} energy cost of enlarging a sufficiently large droplet
has the same leading form in all three regimes. Differentiating
$\Delta E_h(L)$ at large $L$ gives
\begin{equation}
    \frac{d\Delta E_h}{dL}
    \simeq
    \frac{4J_0}{\sigma}L^{-\sigma}-2h .
    \label{eq:dEh_dL}
\end{equation}
The first term is the residual interaction cost of increasing the
domain-wall separation, whereas the second is the bulk energy gained from
converting an additional spin to the stable phase. The critical droplet is obtained when these two contributions balance,
$d\Delta E_h/dL=0$. Its characteristic size is therefore
\begin{equation}
    L_c
    \simeq
    \left(\frac{2J_0}{\sigma h}\right)^{1/\sigma},
    \qquad
    R_c\equiv\frac{L_c}{2}
    \propto h^{-1/\sigma}.
    \label{eq:Rc_lattice}
\end{equation}
Hence the critical-radius exponent varies smoothly across $\sigma=1$,
despite the qualitative change in the total domain-wall energy. 

The activation barrier expressed in terms of the critical
droplet length is
\begin{equation}
    \Delta E_c \simeq
    \begin{cases}
        \dfrac{4J_0}{1-\sigma}L_c^{1-\sigma},
        & 0<\sigma<1, \\[0.8em]
        4J_0\ln L_c+O(1),
        & \sigma=1, \\[0.8em]
        4J_0\zeta(\sigma)
        -\dfrac{4J_0}{\sigma-1}L_c^{1-\sigma},
        & \sigma>1.
    \end{cases}
    \label{eq:lattice_barrier}
\end{equation}
Using Eq.~\eqref{eq:Rc_lattice}, its weak-bias behavior becomes
\begin{equation}
    \Delta E_c(h) \simeq
    \begin{cases}
        B_\sigma h^{(\sigma-1)/\sigma},
        & 0<\sigma<1, \\[0.8em]
        4J_0\ln(2J_0/h)+O(1),
        & \sigma=1, \\[0.8em]
        4J_0\zeta(\sigma)-B_\sigma h^{(\sigma-1)/\sigma},
        & \sigma>1,
    \end{cases}
    \label{eq:lattice_barrier_h}
\end{equation}
where
\begin{equation}
    B_\sigma=\frac{4J_0}{|1-\sigma|}
\Big(\frac{\sigma}{2J_0}\Big)^{1-1/\sigma}>0, \quad \sigma\neq1.
\end{equation}

For low-temperature activated dynamics satisfying detailed balance with the lattice Hamiltonian, this energy barrier controls the leading Arrhenius factor at fixed $h>0$,
$
    \tau_{\rm nuc}\simeq\tau_0\,e^{\Delta E_c/T}.
$
Here $\tau_0$ contains dynamical and entropic contributions and can depend on the bias and system size. At $\sigma=1$ the logarithmic barrier turns the Arrhenius factor into a power law, $\tau_{\rm nuc}\propto h^{-4J_0/T}$, with a temperature-dependent exponent. The result identifies a change in the weak-bias barrier at $\sigma=1$: it diverges as $h\to0^+$ for $\sigma\leq1$ and remains finite for $\sigma>1$. 

\section{Nonlocal field theory}
To obtain a continuum description, we promote the Ising order parameter to a coarse-grained scalar field $\phi(x)$. For interactions decaying as $1/|x-y|^{1+\sigma}$, with $0<\sigma<2$, the corresponding nonlocal quadratic energy can be written as 
\begin{equation}
    \int \mathrm{d}x\, \phi(-\Delta)^{\sigma/2}\phi = \frac{c_\sigma}{2} \int \mathrm{d}x\,\mathrm{d}y\, \frac{[\phi(x)-\phi(y)]^2}{|x-y|^{1+\sigma}}, 
    \label{eq:fractional_quadratic_form} 
\end{equation} 
where $(-\Delta)^{\sigma/2}$ is the fractional Laplacian \cite{kwasnicki2017ten,lischke2020fractional}, with $c_\sigma>0$. In momentum space the Fourier symbol of the fractional Laplacian is $|k|^\sigma$, which dominates over the analytic short-range term $k^2$ at small momentum for all $0<\sigma<2$. We choose units in which the coefficient of the nonlocal term is unity. 

Adding a local potential that supports a metastable and a stable phase gives the coarse-grained free energy 
\begin{equation} 
    F[\phi] = \int \mathrm{d}x\, \left[ \frac{1}{2}\phi(x)(-\Delta)^{\sigma/2}\phi(x) +V(\phi(x)) \right]. \label{eq:Euclidean_action} 
\end{equation} 
We choose a tilted double-well of the form
\begin{equation}
    V(\phi) = \frac{g}{4} (\phi^2-\phi_0^2)^2 - h (\phi+\phi_0),
\label{eq:potential h}
\end{equation}
such that there exists a global stable minimum $\phi_{\rm TV}$, the true vacuum (TV), and a local metastable minimum $\phi_{\rm FV}$, the false vacuum (FV).  
While we determine the minima numerically, for small $h>0$, $\phi_{\rm TV} \approx + \phi_0$ and $\phi_{\rm FV} \approx -\phi_0$.
For the subsequent numerical analysis, we set $g=\phi_0=1$. %

The critical droplet is a nontrivial stationary point of $F[\phi]$ \cite{linde1981fate} and
therefore satisfies
\begin{equation}
    (-\Delta)^{\sigma/2}\phi(x)+V'(\phi(x))=0,
\label{eq:saddle-point}
\end{equation}
where the prime indicates the field derivative.  We seek an even droplet centered at $x=0$, with
\begin{equation}
    \partial_x\phi(0)=0,
    \qquad
    \phi(|x|\to\infty)=\phi_{\rm FV}.
\end{equation}
The relevant escape saddle separates configurations that shrink from those that grow. %
For the candidate critical droplet $\phi_c(x)$, the excess free energy is
\begin{equation}
    \Delta F_c
    =
    F[\phi_c]-F[\phi_{\rm FV}] .
\end{equation}
To leading exponential accuracy, the nucleation time is therefore
\begin{equation}
    \tau_{\rm nuc}
    \simeq
    \tau_0\, e^{\Delta F_c/T}.
\end{equation}
We focus on this leading activation factor and do not evaluate the prefactor $\tau_0$, which involves fluctuation determinants~\cite{PhysRevD.16.1762}.

Before solving the saddle-point equation numerically, which will be described in the next section, we can already find a distinctive consequence of nonlocality by inspecting the asymptotic form of the solution. In fact, it is straightforward to show that if such a critical saddle exists in the present problem, it must develop a power-law tail at long distances.
Let
$\varphi(x)=\phi(x)-\phi_{\rm FV}$ denote the deviation from the metastable
background and define its integrated weight as $M=\int\mathrm{d}y\,\varphi(y)$. Far from the droplet core, $|x-y|\simeq |x|$, the integral representation of the fractional Laplacian therefore gives
\begin{equation}
    (-\Delta)^{\sigma/2}\varphi(x) \simeq -c_\sigma M |x|^{-(1+\sigma)}. 
\end{equation}
Linearizing the saddle-point equation around $\phi_{\rm FV}$, with $m_{\rm FV}^2=V''(\phi_{\rm FV})>0$, then yields
\begin{equation}
    \varphi(x) \simeq \frac{c_\sigma M}{m_{\rm FV}^2} |x|^{-(1+\sigma)}. \label{eq:16}
\end{equation}
The droplet profiles therefore differ qualitatively from those of
a short-range theory: the approach to the metastable background is
algebraic rather than exponential. 
Such algebraic delocalization is a characteristic consequence of long-range interactions and has previously been identified for defect-induced magnetic droplets~\cite{hoyos2007local}. 

\begin{figure}
\centering
\includegraphics[width=1.\linewidth]{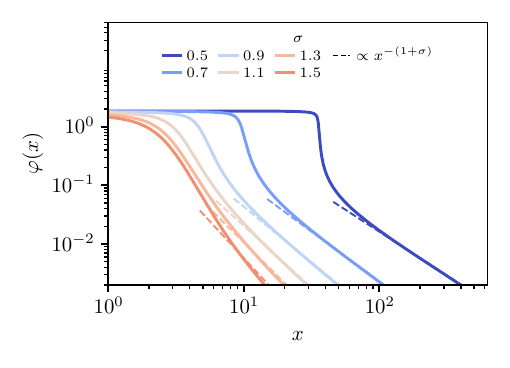}
    \caption{Droplet profile for several
        interaction-range exponents $\sigma$ (color scale) at fixed bias field
        $h = 0.09$. The deviation from the false vacuum $\varphi(x)$ is
        shown versus radius $x$ on log--log axes. Each solid curve is the numerical
        solution and the dashed line of matching color is the predicted power law
        \eqref{eq:16}.}
    \label{fig:algebraic_tail}
\label{fig:solution_tails_sigma}
\end{figure}

\section{Critical droplet}
In this section, we determine the critical domain using two complementary approaches.
First, we compute static saddle-point profiles in the continuum theory described in the previous section. We then locate the threshold
between shrinking and growing domains in stochastic lattice dynamics
and compare the resulting radius scaling with the static prediction.
\subsection{Static determination of the critical droplet}

We first solve the static saddle-point equation \eqref{eq:saddle-point} using the numerical approach
developed in Ref.~\cite{Pagni:2026biq}, where more details on the numerical method can be found. In the present thermal problem, the critical
droplet is static, so we solve only for its spatial profile.
We expand the profile in symmetrized Gaussian functions, evaluate the
fractional Laplacian analytically on this basis, and determine the
expansion coefficients by solving the resulting nonlinear equations.

We obtain nontrivial stationary droplet profiles throughout the range
considered. We choose the computational domain
sufficiently large and verify that the critical radius is insensitive to
further increases in its size.
Solutions are followed across $(\sigma,h)$ by continuation, using a
previously obtained profile for the initial guess at nearby parameters.
As shown in
Fig.~\ref{fig:algebraic_tail}, the numerical tails agree with the asymptotic prediction in Eq.~\eqref{eq:16}.
Smaller values of $\sigma$ yield more slowly decaying tails.
The finite Gaussian expansion approximates this algebraic behavior
over the numerical window, while the asymptotic analysis determines
the decay at arbitrarily large distances.
Equation~\eqref{eq:16} holds for $|x|\gg R_c$. Closer to the droplet, the profile follows the slower tail of a single wall, $\propto|x-R_c|^{-\sigma}$. For the largest $\sigma$ the droplet is small ($R_c\simeq2$ at $\sigma=1.5$), so this crossover falls inside the plotted range. Together with the finite computational domain, it leaves only a short interval in which Eq.~\eqref{eq:16} applies, which explains the visible deviation at $\sigma=1.5$.

We characterize the droplet size by the half-height radius of its
deviation from the metastable background,
$\varphi(R_c)=\varphi(0)/2.
$
As the bias is reduced, the droplet broadens and $R_c$ increases.
The lattice energetics predicts the leading weak-bias behavior
\begin{equation}
    R_c(h)\simeq A_\sigma h^{-1/\sigma},
    \label{eq:Rc_corrections}
\end{equation}
with a nonuniversal amplitude $A_\sigma$ (see
Appendix~\ref{sec:large_radius_scaling} for the corresponding argument carried out directly in the field theory).
\begin{figure*}
    \centering
    \includegraphics[width=0.99\textwidth]{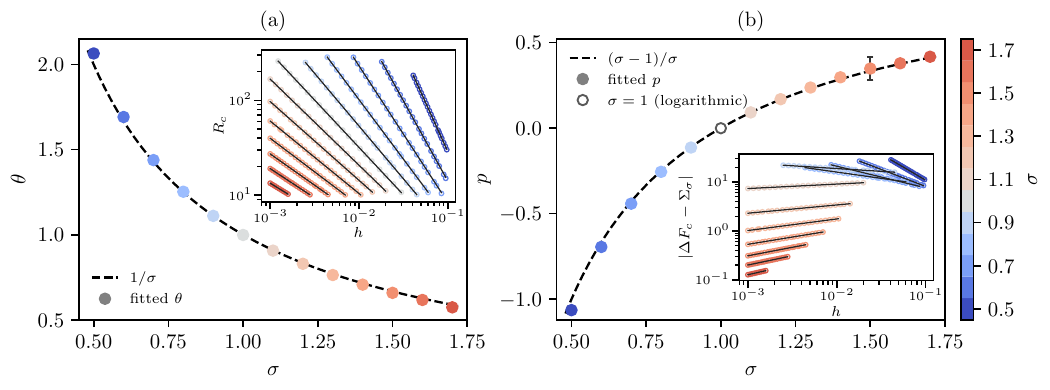}
\caption{Weak-bias scaling of the continuum critical droplet.
(a) Critical-radius exponent $\theta$, obtained from fits to
$R_c\propto h^{-\theta}$, compared with the prediction $1/\sigma$
(dashed line). The inset shows the critical radii $R_c(h)$ (symbols) and their power-law fits (lines).
(b) Barrier exponent $p$, obtained from fits to
$\Delta F_c=\Sigma_\sigma+A'_\sigma h^p$, compared with
$(\sigma-1)/\sigma$ (dashed line). The inset shows the magnitude of the
field-dependent contribution, $|\Delta F_c-\Sigma_\sigma|$ (symbols), with the corresponding fits (lines).
Both insets use logarithmic axes, and the color bar gives $\sigma$ in both panels and insets. All fits use $R_c>10$. Error bars give the change of the exponent when the fit is restricted to the half of these points with the smallest $h$, and are smaller than the symbols except for $p$ at $\sigma=1.5$. %
At $\sigma=1$ (open circle), the same form with free $p$ gives
$p\simeq0$, its logarithmic limit, consistent with
$\Delta F_c\simeq\Sigma_1+a_1\ln(h_*/h)$.}
\label{fig:tau_h}
\end{figure*}

To test this prediction, we fit the numerical radii to
$R_c\propto h^{-\theta}$, treating $\theta$ as a free parameter.
The inset of Fig.~\ref{fig:tau_h}(a) shows the radii and their fits,
while the main panel compares the extracted exponents with
$\theta=1/\sigma$. All fits are restricted to $R_c>10$ to focus on
the large-droplet regime. Over the displayed range, the fitted
exponents agree with the predicted scaling and vary smoothly across
$\sigma=1$.

The activation barrier displays a qualitatively different behavior.
For $\sigma\neq1$, its leading weak-bias form is
\begin{equation}
    \Delta F_c(h)
    \simeq
    \Sigma_\sigma
    +A'_\sigma h^{(\sigma-1)/\sigma},
    \label{eq:deltaFcscaling}
\end{equation}
where $\Sigma_\sigma$ denotes the constant contribution.
For $\sigma>1$, it is the limiting barrier as $h\to0^+$,
whereas for $0<\sigma<1$ it is subleading to the divergent
field-dependent term. At $\sigma=1$, the power law is replaced
by logarithmic growth, see Appendix~\ref{sec:large_radius_scaling}.

We test Eq.~\eqref{eq:deltaFcscaling} by fitting the numerical
barriers to $\Delta F_c=\Sigma_\sigma+A'_\sigma h^p$, with the
offset, amplitude, and exponent treated as free parameters.
The inset of Fig.~\ref{fig:tau_h}(b) shows the magnitude of the barrier after
subtracting the fitted offset.
The main panel compares the extracted exponents directly
with $(\sigma-1)/\sigma$. As for the radius, the fits use only configurations
with large radii ($R_c>10$).
The fitted barrier exponents are consistent with the change
predicted by the lattice energetics: for $0<\sigma<1$, the
field-dependent contribution grows as the bias decreases,
whereas for $\sigma>1$ it vanishes and the barrier approaches
a finite value. The continuum theory thus reproduces the
distinction between a critical radius that diverges on both
sides of $\sigma=1$ and an activation barrier that diverges
only for $\sigma\leq1$. 
At $\sigma=1$, the barrier grows logarithmically. Fitting this case with the
same offset-plus-power-law form gives $p\simeq0$, shown by the open
circle in Fig.~\ref{fig:tau_h}(b). This is consistent with the
logarithmic limit, in which the amplitude and offset compensate
as $p\to0$.

The scalings obtained from the numerical results match the analytical predictions of Sec.~\ref{Sec2:microscopic droplet energetics}, and a similar change occurs at $\sigma=1$ in studies of interfaces in long-range models~\cite{Savin:2012nonlocal}. %

\subsection{Dynamical determination of the critical radius}

As an independent test of the continuum calculation, we determine the
critical radius directly from the microscopic lattice dynamics. We place a
single droplet in the metastable background and find the size that separates
droplets that shrink and disappear from those that grow. We use the
one-dimensional version of the long-range majority-voting dynamics
introduced for a two-dimensional lattice in Ref.~\cite{lazarides2025apparent}.
We prepare a stable-phase domain of width $2R$ inside the
metastable background and evolve the spins through synchronous
updates. At each time step, the probability of assigning
$s_i=+1$ is
\begin{equation}
    p(s_i\to+1)
    =
    \frac{1}{2}
    \left[
        1+\tanh\left(\frac{n_i+h}{T}\right)
    \right],
\end{equation}
with
\begin{equation}
    n_i=
    \frac{1}{\mathcal N_\sigma}
    \sum_{j\neq i}\frac{s_j}{r_{ij}^{1+\sigma}},
    \qquad
    \mathcal N_\sigma=
    \sum_{j\neq i}\frac{1}{r_{ij}^{1+\sigma}},
\end{equation}
where $r_{ij}$ is the periodic distance between sites and
$T$ controls the temperature. %
We use a chain of $N=6000$ sites and fix $T=0.03$. For $\sigma>0$ used in this work, the normalization $\mathcal N_\sigma$
approaches a finite constant as $N\to\infty$.
It therefore fixes the interaction scale without changing
the predicted weak-bias exponent at fixed $\sigma$.
Note that, for fixed parameters of the free-energy functional, the static
critical radius is independent of $T$. Nevertheless,  within the stochastic lattice dynamics, finite
noise can broaden the grow--shrink threshold and shift its
location. We test the predicted radius scaling at fixed low noise and check that it gives temperature-independent results in Appendix~\ref{app:dynamical}.

\begin{figure*}
  \centering
  \includegraphics[width=0.49\textwidth]{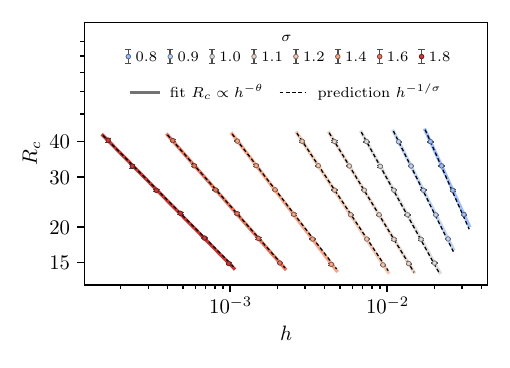}
  \hfill
  \includegraphics[width=0.49\textwidth]{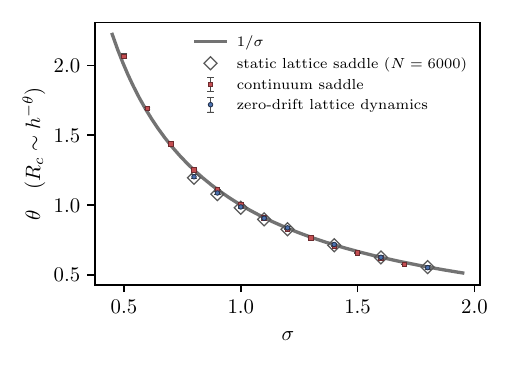}
  \caption{Critical-radius scaling and comparison with static predictions.
Left: dynamical critical radius versus bias at $T=0.03$ and $N=6000$.
Colors denote the interaction exponent $\sigma$. Circles show
the zero-drift estimates. Error bars combine statistical uncertainties
with a $0.5\%$ systematic contribution. Solid lines are weighted fits over all fields
to $R_c=A_\sigma^{\rm dyn}h^{-\theta}$, and dashed lines show
$h^{-1/\sigma}$. Both axes are logarithmic.
Right: fitted exponents from the continuum saddle profiles
(red squares), zero-drift lattice dynamics (blue circles), and
static lattice radii (diamonds). The lattice fits use the fields with $R_c^{\rm st}\ge20$, as in Table~\ref{tab:theta_dyn}, and the static fits the same chain length and fields as the dynamical ones.
The gray curve denotes $1/\sigma$.}
  \label{fig:Rc_scaling}
  \label{fig:appB_Rch}
  \label{fig:lattice_radius_fits}
\end{figure*}

At fixed parameters $(\sigma,h,T)$ we prepare domains of different initial sizes and
measure their mean growth rate over a short time window. We define the
dynamical critical radius as the size at which this mean rate of growth vanishes. %
For each value of $h$ we use 24 independent seed sets of
$1.1\times10^{4}$ trajectories each, 
and the error bars combine statistical uncertainties with a $0.5\%$ systematic contribution.  The details on the procedure
can be found in Appendix~\ref{app:dynamical}. The extracted radii satisfy
$R_c\ll N$ throughout the range considered. For each interaction exponent, we
vary the bias and fit
$
  R_c(h) = A^{\rm dyn}_\sigma h^{-\theta},
$
with both the amplitude $A^{\rm dyn}_\sigma$ and the exponent $\theta$ treated
as fit parameters. The left panel of Fig.~\ref{fig:Rc_scaling} shows the radii and
fits over all fields, whose static critical radii lie in $R_c^{\rm st}=15$--$40$. The exponents in its right panel and in Table~\ref{tab:theta_dyn} use only the fields with
$R_c^{\rm st}\ge20$ (Appendix~\ref{app:finite_chain}). All radii lie on the fitted power laws
within the quoted uncertainties.
The right panel of Fig.~\ref{fig:Rc_scaling} compares the resulting exponents
with those extracted from the continuum saddle profiles
and with the analytical prediction $\theta=1/\sigma$.
The two independent determinations closely follow the
predicted dependence on the interaction range, including
across $\sigma=1$.

\section{Conclusions and Outlook}
\label{sec:conclusions}

We have investigated how long-range interactions modify thermal
droplet nucleation in one-dimensional systems. Unlike in
the nearest-neighbor chain, where the energy of a domain-wall pair is
independent of its separation, power-law interactions introduce
a size-dependent energetic cost that competes with the bulk
free-energy gain. Combining microscopic lattice energetics
with saddle-point solutions of a nonlocal $\phi^4$ field theory,
we have established the weak-bias scaling of the critical
droplet radius and activation barrier.

Our central result is that the divergence of the activation barrier
as $h\to0^+$ switches on at $\sigma=1$.
While the critical radius follows $R_c\propto h^{-1/\sigma}$
throughout the interaction range considered, the barrier
exhibits three regimes. For $0<\sigma<1$, it diverges
algebraically as $h\to0^+$; at $\sigma=1$, the divergence
becomes logarithmic; and for $\sigma>1$, the barrier
approaches a finite value. The threshold $\sigma=1$ thus
separates qualitatively different metastability regimes
and coincides with the interaction-range boundary for
finite-temperature ordering in the long-range Ising
chain~\cite{Dyson:1968up,thouless1969long,frohlich1982phase}. For $\sigma>1$, where the chain has no ordered phase at $T>0$, the single-droplet picture as $h\to0^+$ therefore requires $T\ll4J_0\zeta(\sigma)$ and a finite system size or observation time, before thermally excited domain walls proliferate.

The continuum calculation further resolves the spatial
structure of the critical droplets, revealing algebraic
tails characteristic of nonlocal interactions. Independently,
stochastic long-range lattice dynamics identifies the critical
radius through the threshold between shrinking and growing
domains. The resulting scaling exponents agree with the
analytical prediction and closely track those obtained from
the corresponding static lattice problem. By examining
finite-chain and temperature effects, we show that the
observed deviations from the asymptotic scaling are largely
accounted for by finite-size corrections.

The activation barrier determines the leading exponential
dependence of the thermal nucleation time in the weak-noise
regime. A complete description of metastable decay, however,
also requires the nucleation prefactor and the subsequent
evolution of the emerging domains. Extending the present
framework to nucleation
precursors~\cite{Pirvu:2023plk,Pirvu:2024nbe}
and post-nucleation growth~\cite{bray1993domain}
would clarify how long-range interactions influence the full
decay process. In particular, interactions between spatially
separated droplets may modify relaxation
timescales~\cite{Worm_2013} and enable cooperative decay
mechanisms~\cite{mccraw1980metastability,
chandra1989nucleation,PhysRevE.54.1383},
beyond the single-droplet picture considered here.

\begin{acknowledgments}
    We thank M. Delladio, T. Donner, S. Floerchinger, G. Gori, E. Grossi, G. Morigi and A. Solfanelli for useful discussions.  This research was funded by the Swiss State Secretariat for Education, Research and Innovation (SERI), Swiss National Science Foundation (SNSF) grant numbers 200021--207537, 200021--236722,  and by NCCR SwissMAP. 
\end{acknowledgments}

\appendix
\begingroup
\allowdisplaybreaks

\section{Large-radius scaling of the critical droplet}
\label{sec:large_radius_scaling}

In this appendix we derive the scalings of the radius and the barrier [Eqs.~\eqref{eq:Rc_corrections} and~\eqref{eq:deltaFcscaling}, respectively].
We consider a potential term in the form~\eqref{eq:potential h}
and denote the free-energy-density difference between the false and true vacua by
\begin{equation}
    \epsilon(h)=V(\phi_{\rm FV})-V(\phi_{\rm TV})
    =2\phi_0 h+\mathcal O(h^3).
\end{equation}
For a large droplet of diameter $2R$, the leading radius-dependent free energy (for $\sigma\neq1$) has the form
\begin{equation}
    \Delta F_0(R,h)
    =\Sigma_0+a_\sigma R^{1-\sigma}-2\epsilon( h) R.
    \label{eq:largeR_leading_energy}
\end{equation}
 It is useful to define the positive long-range force amplitude
$
    \kappa_\sigma\equiv(1-\sigma)a_\sigma>0.
$
For $\sigma>1$, this simply implies $a_\sigma<0$: the interaction energy approaches its asymptotic constant from below, while its derivative with respect to $R$ remains positive.

The critical droplet is the saddle with respect to its size.  Stationarity implies
\begin{equation}
   \partial_R\Delta F_0=0 \implies 2\epsilon=\kappa_\sigma R_c^{-\sigma},
\end{equation}
so that
\begin{equation}
    R_c=A_\sigma h^{-1/\sigma},
    \label{eq:Rc_leading}
\end{equation}
where $
    A_\sigma=\left[\kappa_\sigma/(4 \phi_0)\right]^{1/\sigma}$.

Using the stationarity condition once more, the critical free energy becomes
\begin{equation}
    \Delta F_{0,c}
    =\Sigma_\sigma+A'_\sigma h^{(\sigma-1)/\sigma},
    \label{eq:Fc_leading}
\end{equation}
where $\Sigma_\sigma\equiv\Sigma_0$ and $A'_\sigma=\sigma a_\sigma A_\sigma^{1-\sigma}$, as in Eq.~\eqref{eq:deltaFcscaling}.  

At $\sigma=1$, the leading interfacial contribution grows
logarithmically with the droplet radius. The large-radius
free energy therefore takes the form
\begin{equation}
    \Delta F_0(R,h)
    = \Sigma_0 + a_1 \ln(R/\ell)
    - 2\epsilon(h)R,
\end{equation}
where $a_1>0$ and $\ell$ is a fixed microscopic length. %
The stationarity condition gives
\begin{equation}
    R_c=\frac{a_1}{2\epsilon(h)}
    \simeq \frac{a_1}{4\phi_0 h}.
\end{equation}
Thus, the leading radius scaling remains $R_c\propto h^{-1}$,
consistent with the general prediction $R_c\propto h^{-1/\sigma}$.
Evaluating the free energy at this radius yields
\begin{equation}
    \Delta F_{0,c}
    = \Sigma_0
    + a_1 \ln\!\left[\frac{a_1}{2\epsilon(h)\ell}\right]
    - a_1.
\end{equation}
Absorbing the field-independent terms into a constant $\Sigma_1$,
the leading weak-bias behavior can be written as
\begin{equation}
    \Delta F_c(h)
    = \Sigma_1+a_1\ln(h_*/h)+o(1),
    \qquad h\to0^+,
\end{equation}
with $h_*>0$ a fixed reference bias. The logarithm is therefore
the leading divergence of the barrier at $\sigma=1$.

\section{Extraction of the scaling exponent}
\label{app:dynamical}

In this appendix we explain how we determine the dynamical critical radius and extract
its dependence on the bias $h$. We also examine the effects of the
observation time and temperature. 

\subsection{Prepared domains and the static reference}

We use the synchronous spin dynamics described in the main text on a periodic
chain of $N=6000$ sites. The interaction between two distinct sites is
$1/(\mathcal N_\sigma r_{ij}^{1+\sigma})$, where $r_{ij}$ is their periodic
distance and $\mathcal N_\sigma$ is the normalization used in the update
rule. %
Unless stated otherwise, the noise parameter is $T=0.03$. Starting from the metastable background $s_i=-1$, we reverse a contiguous
interval of $D_0$ spins. During the evolution we record
\begin{equation}
  D(t)=\sum_i\frac{1+s_i(t)}{2}.
  \label{eq:appB_domain_size}
\end{equation}
For a single contiguous domain, this is its width, so that $R=D/2$.
The excess energy of a
contiguous domain $\mathcal D$ containing $D$ up spins is
\begin{equation}
  \Delta E_h(D)
  =2\sum_{i\in\mathcal D}\sum_{j\notin\mathcal D}
    \frac{1}{\mathcal N_\sigma r_{ij}^{1+\sigma}}-2hD.
  \label{eq:lattice_energy}
\end{equation}
 Their
competition produces a maximum at $D_c^{\rm st}$, defining
$R_c^{\rm st}=D_c^{\rm st}/2$.

\begin{figure}
  \centering
  \includegraphics{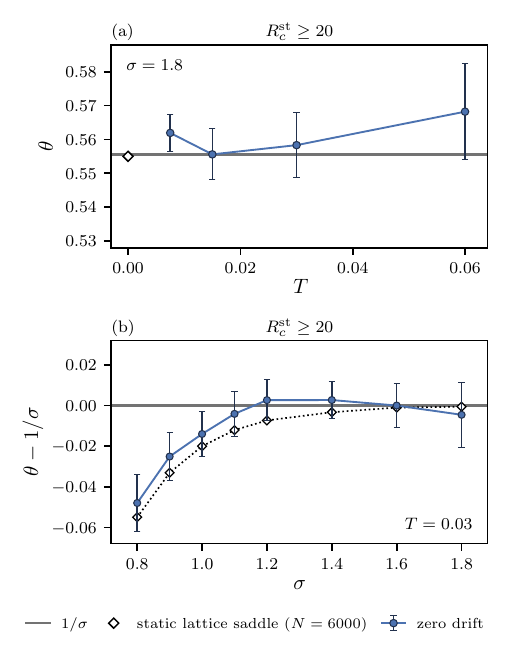}
\caption{Zero-drift exponents fitted using only fields with
$R_c^{\rm st}\ge20$.
(a) Temperature dependence at $\sigma=1.8$. Filled circles show the
exponents extracted at $T=0.0075$, 0.015, 0.03, and 0.06. The horizontal
line denotes $1/1.8$, and the diamond is the static lattice reference.
(b) Deviation $\theta-1/\sigma$ versus $\sigma$ at $T=0.03$.
Blue circles show the zero-drift estimates, and diamonds the static
lattice exponents at $N=6000$, fitted over the same fields.
The horizontal line denotes zero deviation from $1/\sigma$. The downward shift at small $\sigma$ is a finite-chain effect, see
Appendix~\ref{app:finite_chain}. Error bars are obtained from the covariance
of the weighted fits.}
  \label{fig:appB_T}
\end{figure}
\begin{figure}
	\centering
  \includegraphics{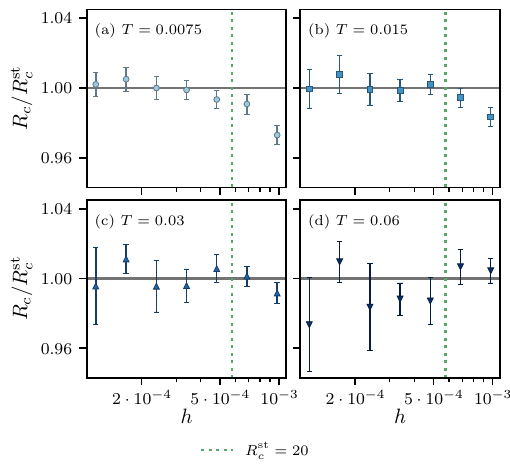}
	\caption{Ratio of the zero-drift critical radius to the static lattice radius at $\sigma=1.8$ for (a)~$T=0.0075$, (b)~0.015, (c)~0.03, and (d)~0.06.   Vertical 
    dotted lines mark $R_c^{\text st}=20$. }
	\label{fig:appB_fitsT}
\end{figure}

\subsection{Critical threshold}

To characterize the radius numerically, we use as a criterion the size at which the mean motion of the walls changes direction. In practice, after discarding the first $\tau_1=10$
updates to allow the initially sharp walls to relax, we measure the
conditional mean change in width,
\begin{equation}
  v(D;\tau_2)
  =\big\langle D(\tau_2)-D(\tau_1)
    \,\big|\,D(\tau_1)=D\big\rangle.
  \label{eq:drift}
\end{equation}
Dividing by $\tau_2-\tau_1$ gives the mean growth rate. We define
\begin{equation}
  v(D^\ast;\tau_2)=0,
  \qquad R_c(\tau_2)=\frac{D^\ast(\tau_2)}{2}.
  \label{eq:appB_zero_drift}
\end{equation}
Below this size the mean change is negative; above it the mean change is
positive. Averaging
many trajectories allows us to resolve the change of sign even when the
motion of a single domain is dominated by noise.

At each $(\sigma,h)$, we prepare eleven initial widths uniformly spanning
$2R_c^{\rm st}[0.75,1.25]$.
For each width we run $10^3$ independent
trajectories in each of 24 independent seed sets. We record the widths at $\tau_1$ and at
$
  \tau_2\in\{11,12,14,16,20,25,30,40,50,60,80\}.
$
Following many domains, we record their change in width
$D(\tau_2)-D(\tau_1)$ as a function of the initial width $D(\tau_1)$.
The critical width $D^\ast$ is where shrinking turns into growing on
average, which we locate as the zero of a quadratic fit to these data,
and $R_c=D^\ast/2$.
The fitted
zero closest to the median sampled width gives $D^\ast(\tau_2)$. %
Since domains change size slightly during the first $\tau_1$ steps, we use
their width at $\tau_1$, when the measurement starts, rather than the
initial width $D_0$.
The observation window must be long enough to resolve the drift, but short
enough that a domain does not move far from its starting width. We therefore
look for a range of $\tau_2$ over which $R_c(\tau_2)$ is stable.  For each $h$, the estimate of $R_c$ depends on the window length $\tau_2$ when the window is too short or too long, and is constant in between. Among sequences of at least three consecutive measurement times whose radius
estimates agree with their mean within two standard errors, we select the
sequence with the smallest uncertainty on that mean, and report the mean over
this plateau as $R_c$. Statistical uncertainties are estimated from six groups
of four seed sets (six groups of eight for the fields measured with 48 seed
sets, see below). A systematic uncertainty of $0.5\%$ of the radius is added
in quadrature for the choice of time window. Fields without a plateau are excluded. This removes the smallest domains for $\sigma=0.8$ and $0.9$,
leaving four and five fields, respectively, and six for the other interaction
exponents.

For each interaction exponent we fit the accepted radii to
$
  \ln R_c=\ln A_\sigma^{\rm dyn}-\theta\ln h,
$
where the amplitude and exponent are free parameters in a weighted
least-squares fit. The uncertainty on $\theta$ is obtained from the fit
covariance. The left panel of Fig.~\ref{fig:Rc_scaling} shows that a power law describes
the sampled field range. We examine the deviations from the predicted
exponent $1/\sigma$ below.

\subsection{Finite-chain effects and the fitting range}
\label{app:finite_chain}

Table~\ref{tab:theta_dyn} compares the dynamical exponent with the exponent
obtained by fitting the static lattice radii.
Using the same chain and fit range separates deviations already present
in the static problem from those introduced by the dynamics or its
measurement.

\begin{table}[t]
  \caption{Exponents fitted from the zero-drift radii at $T=0.03$ and
  from the static lattice radii at the same $N=6000$ and over the same
  retained fields, $R_c^{\rm st}\ge20$, as in Fig.~\ref{fig:appB_T}(b). The final column is the asymptotic prediction.}
  \label{tab:theta_dyn}
  \begin{ruledtabular}
  \begin{tabular}{cccc}
  $\sigma$ & $\theta_{\rm dyn}$ & $\theta_{\rm st}$ & $1/\sigma$ \\
  \hline
  0.8 & $1.202\pm0.014$ & 1.195 & 1.250 \\
  0.9 & $1.086\pm0.012$ & 1.078 & 1.111 \\
  1.0 & $0.986\pm0.011$ & 0.980 & 1.000 \\
  1.1 & $0.905\pm0.011$ & 0.897 & 0.909 \\
  1.2 & $0.836\pm0.010$ & 0.826 & 0.833 \\
  1.4 & $0.717\pm0.009$ & 0.711 & 0.714 \\
  1.6 & $0.625\pm0.011$ & 0.624 & 0.625 \\
  1.8 & $0.551\pm0.016$ & 0.555 & 0.556 \\
  \end{tabular}
  \end{ruledtabular}
\end{table}

For $\sigma\lesssim1.1$, the static exponent itself lies below
$1/\sigma$. The finite periodic chain omits interactions beyond $N/2$,
changing the attractive interaction between the walls by a contribution of order
$(N/2)^{-\sigma}$. For domains much smaller than the chain, this acts
approximately as a small shift of the bias. Its relative importance increases
as $h$ decreases and is more pronounced for slowly decaying interactions.
Consequently, choosing $R_c\ll N$ alone does not eliminate the finite-chain corrections
to the fitted exponent. Repeating the static calculation on longer chains
brings the exponent closer to $1/\sigma$: at $\sigma=0.8$, over the same fields,
$\theta_{\rm st}=1.195$, $1.241$ and $1.249$ for $N=6\times10^3$, $6\times10^4$
and $6\times10^5$, against $1/\sigma=1.25$.

The dynamical estimates follow this finite-chain trend. Residual
differences from the static radii are small for domains with
$R_c\gtrsim20$, where the zero-drift and static radii agree to within
$1\%$. The largest deviations, about $2\%$, occur for the smallest
domains ($R_c^{\rm st}=15$) at $\sigma=1.2$ and $1.4$, and such deviations can tilt the power-law fit. This motivates restricting the exponent fits below to fields with
$R_c^{\rm st}\ge20$.

Figure~\ref{fig:appB_T}(b) compares the deviations of the static
and zero-drift exponents from $1/\sigma$ after imposing this cut.
The static results identify the shift already present in the
finite-chain problem, while their difference from the dynamical
results indicates the additional effect of the dynamics and its
measurement. The remaining downward shift at small $\sigma$ shows
that excluding small domains does not remove finite-chain effects.

\subsection{Dependence on the noise strength}

We test the dependence on the noise strength at a chosen $\sigma=1.8$ using
$T=0.0075$, $0.015$, $0.03$, and $0.06$. Since the domain energy in
Eq.~\eqref{eq:lattice_energy} has no explicit $T$ dependence, its maximum
provides a fixed reference for this comparison.
At the two largest temperatures $T=0.03$ and $0.06$, we use 48 seed sets and
extend the measurement times to $\tau_2=100$, $120$, and $160$, since the
drift of large domains is slower.

For $T=0.03$ and $\sigma=1.8$, Fig.~\ref{fig:appB_T}(a) uses this extended
data set, which includes a longer measurement of the $R_c^{\rm st}=40$ field,
whereas Fig.~\ref{fig:appB_T}(b) uses the production fields of Table~\ref{tab:theta_dyn}. The resulting exponents agree within their
uncertainties.

Figure~\ref{fig:appB_fitsT} shows the ratio of the measured radii to the static
reference.
At low temperature, the smallest domains, corresponding to the largest
biases, lie below the static radius. These deviations are consistent
with the discrete motion of the walls and steepen the fitted power law.
At higher temperature, the deviations are less systematic, while larger
uncertainties reflect the difficulty of resolving a weak mean drift
amid width fluctuations.
Figure~\ref{fig:appB_T}(a) shows the fitted exponents at these four
temperatures using only fields with $R_c^{\rm st}\ge20$.
Within the quoted uncertainties, the results are consistent with a
temperature-independent exponent and with $1/1.8\simeq0.556$.

\endgroup

\bibliography{ref_lib_classical_inspire}

\end{document}